\documentclass[11pt]{article}

\usepackage[final]{acl}

\usepackage{times}
\usepackage{latexsym}

\usepackage[T1]{fontenc}

\usepackage[utf8]{inputenc}
\usepackage{amsmath}
\usepackage{amsfonts}
\usepackage{amssymb}
\usepackage{multirow}
\usepackage{tabularx}
\usepackage{mathtools}
\usepackage{booktabs}
\usepackage{float}
\usepackage{caption} 
\usepackage{needspace} 
\usepackage{enumitem}
\usepackage[most]{tcolorbox}
\usepackage{microtype}

\usepackage{inconsolata}

\usepackage{graphicx}

\title{%
  \raisebox{1cm}[0pt][0pt]{%
    \makebox[\textwidth][c]{%
      \scalebox{1}[1]{%
        \includegraphics[height=0.88cm,trim=0 4 0 4,clip]{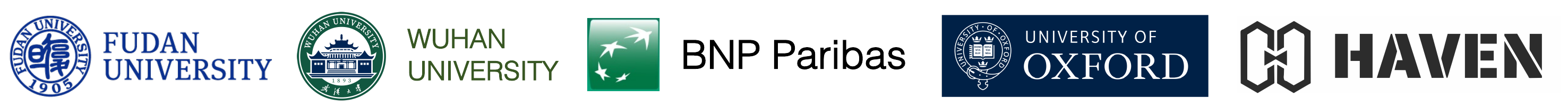}%
      }%
    }%
  }\\[-2.8em]
  \rule{\textwidth}{0.4pt}\\[1.2em]
  Behavioral Consistency Validation for LLM Agents: An Analysis of Trading-Style Switching through Stock-Market Simulation
}

\author{
  Zeping Li$^{1}$,
  Guancheng Wan$^{2}$,
  Keyang Chen$^{1}$,
  Yu Chen$^{3}$,
  Yiwen Zhao$^{1}$, \\
  \textbf{Philip Torr}$^{4}$,
  \textbf{Guangnan Ye}$^{1}$,
  \textbf{Zhenfei Yin}$^{4,5,\dagger}$
  \textbf{Hongfeng Chai}$^{1,\dagger}$, \\
  $^{1}$Fudan University \quad
  $^{2}$Wuhan University \quad
  $^{3}$BNP Paribas \quad
  $^{4}$Oxford University \\ 
  $^{5}$Haven
  $^{\dagger}$Corresponding authors
}

\begin{document}
\maketitle
\begin{abstract}
Recent works have increasingly applied Large Language Models (LLMs) as agents in financial stock market simulations to test if micro-level behaviors aggregate into macro-level phenomena. However, a crucial question arises: Do LLM agents' behaviors align with real market participants? This alignment is key to the validity of simulation results. To explore this, we select a financial stock market scenario to test behavioral consistency. Investors are typically classified as fundamental or technical traders, but most simulations fix strategies at initialization, failing to reflect real-world trading dynamics. In this work, we assess whether agents’ strategy switching aligns with financial theory, providing a framework for this evaluation. We operationalize four behavioral-finance drivers—loss aversion, herding, wealth differentiation, and price misalignment—as personality traits set via prompting and stored long-term. In year-long simulations, agents process daily price-volume data, trade under a designated style, and reassess their strategy every 10 trading days. We introduce four alignment metrics and use Mann–Whitney U tests to compare agents’ style-switching behavior with financial theory. Our results show that recent LLMs' switching behavior is only partially consistent with behavioral-finance theories, highlighting the need for further refinement in aligning agent behavior with financial theory.
\end{abstract}

\section{Introduction}
Agent-Based Modeling (ABM) has a long tradition in economics and the social sciences for explaining how micro-level behavioral rules generate macro-level phenomena ~\cite{schelling,epstein,bonabeau}. Classic ABMs typically rely on handcrafted heuristics and fixed rule sets, which improve interpretability yet, by construction, restrict agents’ perceptual and decision spaces, thereby constraining system heterogeneity. Recent progress in LLMs offers an alternative way to instantiate agents in ABM systems ~\cite{LLM-ABM-survey1,lu2024llms}. LLMs are capable of processing unstructured and structured signals, reasoning over extended context, and producing explicit rationales for actions ~\cite{zhao2023survey}. Through prompting, LLM-based agents can be configured to express diverse personality traits, expanding agent-level heterogeneity in roles and decision styles ~\cite{shao2023character,li2023camel}. Incorporating LLMs as agents thus yields simulations that more closely approximate complex real-world systems. As a result, LLM-enabled ABM is now increasingly used for simulation across complex social systems, game-theoretic environments, and financial markets ~\cite{park2023generative,kovavrik2025game,zhang2024multimodal}.

In financial markets, there has been a substantial body of work on simulations using LLM agents~\cite{wu2023bloomberggpt,lai2024sec,LLM-ABM-stock-market1-twinmarket,LLM-ABM-stock-market2-cantrade}, where these agents interpret financial news and corporate disclosures to generate trading strategies based on price and volume histories. These strategies are often converted into standard fundamental and technical indicators~\cite{xiao2024tradingagents}. However, these studies typically fix investors as either fundamental or technical traders at initialization, with no strategy switching occurring during the simulation. This limitation contrasts with traditional financial studies~\cite{behavioral-drivers3-structural}, which show that a larger proportion of technical traders amplifies boom-bust cycles, while a greater share of fundamental traders stabilizes prices. Thus, explicit style switching is crucial to reproduce stylized facts such as volatility clustering, long-range dependence, and heavy-tailed returns. Before incorporating style-switching behavior into simulations, we pose a fundamental and critical question: under realistic information and constraints, can the style-switching behavior of LLM agents align with that of real market participants?

To this end, we examine the behavioral consistency of LLM agents in their style-switching behavior. Drawing from four key behavioral finance drivers—Loss aversion tendency~\cite{behavioral-drivers4-lossaversion}, Herding tendency~\cite{behavioral-drivers5-herding}, Wealth differentiation sensitivity~\cite{behavioral-drivers6-differential}, and Price misalignment sensitivity~\cite{behavioral-drivers7-misalignment}—we map these factors to agent-level personality traits through prompting and embed them as long-term memories that guide strategy formation and style-switching decisions. The evaluation runs on a stock-market simulator using S\&P 500 stocks in 2024, combining daily price–volume data with curated trading indicators. Each round, agents operate under a designated style, and a counterfactual ledger tracks the alternative; every ten trading days, agents review returns and traits to decide whether to switch styles.

To assess the behavioral fidelity of LLM agents, we pose four research questions aligned with behavioral finance theories: \textbf{(RQ1)} Does loss aversion keep agents in their current style after losses? \textbf{(RQ2)} Does herding increase switching when the alternative style has a larger population share? \textbf{(RQ3)} Does wealth differentiation trigger switching when the alternative style outperforms? \textbf{(RQ4)} Does price misalignment induce shifts toward the fundamental style when market prices diverge from estimated value? We validate these questions by comparing agents with aligned and non-aligned traits using Mann–Whitney U tests. The results show that LLM agents' switching behavior is only partially consistent with behavioral-finance theories. In summary, the main contributions of this paper are as follows:
\begin{itemize}
    \item To our knowledge, we are the first to identify style switching in LLM agents within market simulations. We consider four behavioral drivers that influence style transitions and embed them into the agents' long-term memory.
    \item We propose four alignment metrics and use Mann–Whitney U tests to compare aligned and non-aligned cohorts, providing an operational test for the consistency of LLM agents’ style-switching behavior with financial theory.
    \item Year-long simulations conducted with various recent LLM agents show that while LLM agents exhibit partial consistency with behavioral-finance theories, they cannot fully align with these theories in all aspects.
\end{itemize}

\section{Related Works}
\subsection{Traditional ABM Simulation}
Traditional agent-based models have been used in finance for decades to link heterogeneous trading rules with price formation and market statistics. Early artificial–stock-market work shows that adaptive, learning agents in centralized venues generate realistic return dynamics and position updates ~\cite{traditional-ABM1}, with subsequent formulations of endogenous expectations closing the loop from beliefs to orders and prices ~\cite{traditional-ABM2}. Chartist–fundamentalist multi-agent models reproduce stylized facts such as fat tails and clustered volatility ~\cite{traditional-ABM3}, while mechanism studies connect executable rules (value, trend following, market making) to volatility and liquidity ~\cite{traditional-ABM4}; microsimulations that embed heterogeneity, peer interactions, and trade frictions account for serial-dependence patterns ~\cite{traditional-ABM5}. Taken together, ABMs’ explicit heterogeneity, interaction, and microstructure realism make them well suited to simulate financial markets end-to-end and to validate how micro behavior aggregates into macro phenomena ~\cite{ABM-survey1}.

\subsection{LLM-Based ABM Simulation}
Recent advances in LLMs have rapidly increased the use of LLM-enabled ABM in finance and economics, offering more realistic and scalable simulations than traditional systems~\cite{LLM-ABM-survey1,LLM-ABM-survey2}. This progress is driven by three core capabilities of LLM agents: (i) processing rich textual information with price–volume signals, (ii) maintaining persona-like memory to sustain heterogeneous behavior, and (iii) emitting structured actions (e.g., function-call orders) integrated with market microstructure. Building on this, ~\cite{LLM-ABM-fin-eco-simulation1} embeds LLM agents in ABM economies to produce interpretable, heterogeneous multi-period behavior, while ~\cite{LLM-ABM-fin-eco-simulation2} scales to thousands of agents to replicate macro expectation formation (inflation, unemployment).

In financial stock markets, ~\citet{LLM-ABM-stock-market1-twinmarket} integrates a networked communication layer within a BDI-style framework, tracing information propagation through trading decisions to emergent macro phenomena; it reproduces mechanisms like bubbles and drawdowns, enabling scalable behavioral–social simulations. Similarly, ~\cite{LLM-ABM-stock-market2-cantrade} builds a continuous double auction with function-call decisions, and ~\cite{LLM-ABM-stock-market3-stockagent} and ~\cite{LLM-ABM-stock-market4-gao} provide multi-agent trading in realistic environments. ~\cite{LLM-ABM-stock-market5-fincon} enhances financial decision-making, and ~\cite{LLM-ABM-stock-market6-shiftingpower} explores controllable agent heterogeneity through preference-based prompts (e.g., risk or ambiguity aversion). However, a recent study ~\cite{LLM-ABM-stock-market7-henning} finds that markets populated by LLM agents often appear "too rational" compared to human-subject experiments, raising doubts about their behavioral fidelity in stock-market simulations. This paper thus focuses on testing whether LLM agents' style-switching behavior aligns with financial theory.

\begin{figure*}[t]
  \centering
  \includegraphics[width=\textwidth]{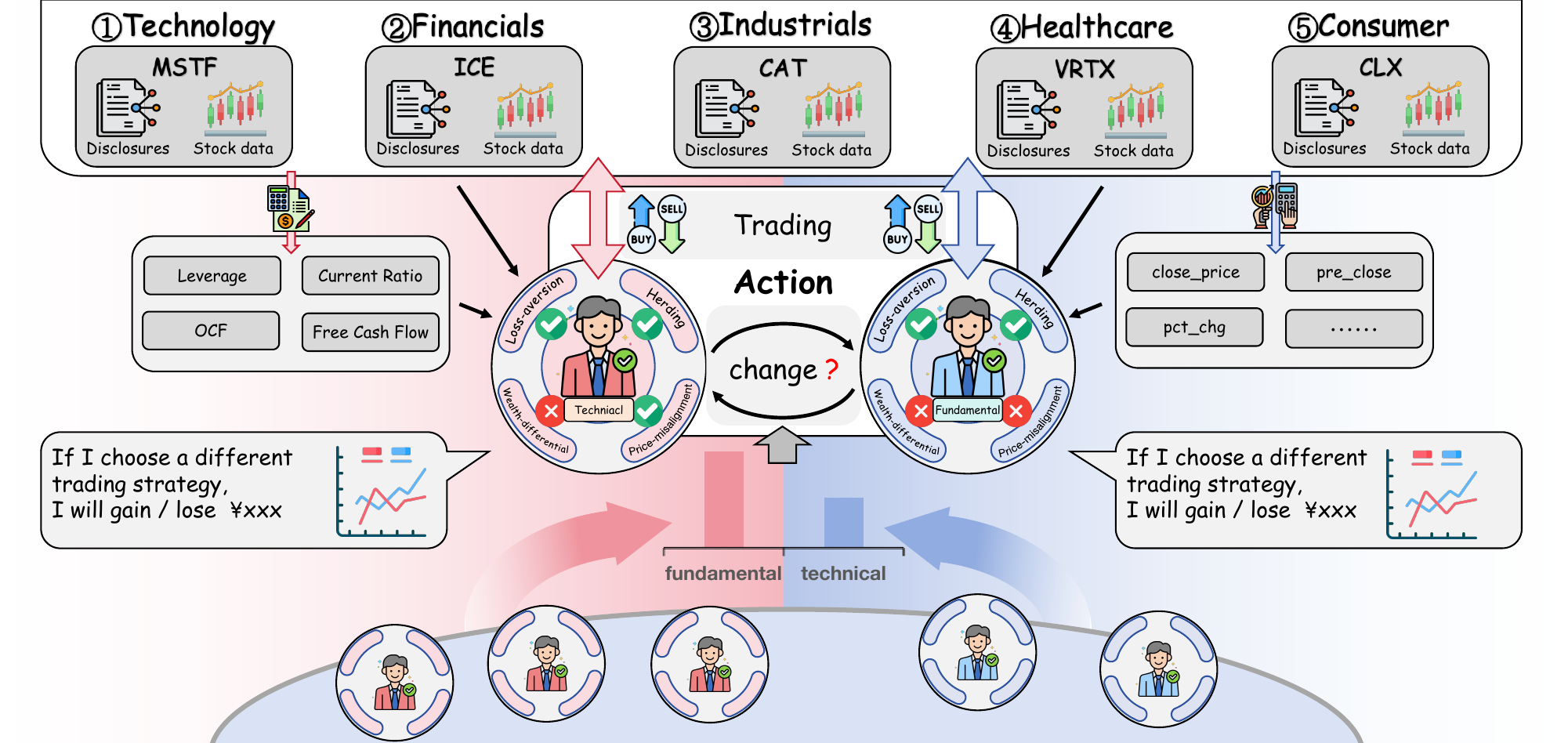}
  \caption{Overview of the simulation framework. Agents trade under a designated technical or fundamental style, using daily price–volume data and quarterly disclosures to compute indicators. Each agent carries different behavioral drivers—loss aversion tendency, herding tendency, wealth differentiation sensitivity, and price misalignment sensitivity—that shape trading decisions and periodic style switching, thereby updating population shares.}
  \label{fig:logo}
\end{figure*}

\subsection{Behavioral Drivers of Style Switching}
To understand style switching, we start with ~\cite{behavioral-drivers1-Brock,behavioral-drivers2-Brock}, who model agents’ switches between forecasting rules (e.g., fundamentalist versus trend-following) via a discrete-choice mechanism based on relative performance, providing a foundation for endogenous switching. Extending this line of work, ~\citet{behavioral-drivers3-structural} model switching between technical and fundamental traders via a relative ``attractiveness'' index that aggregates predisposition, herding, mispricing, and wealth differentiation, thereby supporting the model's reproduction of key financial stylized facts. 

Guided by these works, we select four theory-grounded factors that map directly onto our research questions. Specifically, loss aversion ~\cite{behavioral-drivers4-lossaversion} predicts persistence in the current style even after recent losses (RQ1); herding ~\cite{behavioral-drivers5-herding} implies a higher likelihood of switching into the majority style when population shares tilt toward it (RQ2); wealth differentiation ~\cite{behavioral-drivers6-differential} suggest switching toward the counterfactual style as the counterfactual style’s recent outperformance widens (RQ3); and price misalignment ~\cite{behavioral-drivers7-misalignment} predicts movement toward the fundamental style as the gap between market price and fundamental value grows (RQ4). This mapping transforms classic behavioral-finance mechanisms into testable predictions about when and how LLM agents should switch styles. By modeling style-switching behavior around these key behavioral drivers, we aim to address the gap in the literature regarding LLM agents' ability to replicate such behavior, particularly their capacity to switch strategies in response to these cues.

\section{Method}%
\label{sec:method}%

\subsection{Stock Market Simulation Setup} 
\noindent\textbf{Overall setup.} We collect a full-year 2024 dataset for the S\&P 500 constituents, integrating two modalities per ticker: (i) daily price--volume records with dividend and split flags; (ii) quarterly corporate disclosures---balance sheet and cash-flow statement. The price–volume stream provides the primitives for standard technical features used by technical traders. Quarterly balance sheet and cash-flow statement disclosures supply the structured inputs used to construct solvency, liquidity, and cash-generation metrics for fundamental analysis. The concrete fields required for feature construction are summarized in Table~\ref{table-data-fields}. More implementation details can be found in Appendix~\ref{implementation_details}.

\noindent\textbf{Technical indicators.} From the fields in Table~\ref{table-data-fields}, we derive the technical indicators listed in Table~\ref{table-indicators} , including daily change, percentage change, volume, and moving averages, which capture short- and long-horizon trends and liquidity.

\noindent\textbf{Fundamental indicators.} We retain four key metrics (Table~\ref{table-indicators}): (i) Leverage for solvency risk; (ii) Current Ratio for liquidity; (iii) OCF for cash from operations; and (iv) FCF for discretionary cash flow. These cover the key axes of solvency, liquidity, and cash generation.

\noindent\textbf{Stock pool construction.}
We select one high-quality representative from each of five stock sectors—Information Technology, Financials, Health Care, Industrials, and Consumer Staples—forming a simulation pool with MSFT, ICE, VRTX, CAT, and CLX. Details of the construction can be found in the Appendix~\ref{stock_pool_details}.

\subsection{Heterogeneous Agent Initialization}
\noindent\textbf{Factorial Design of Agent Heterogeneity.}  
We instantiate a balanced population of $2^5=32$ agents using a full-factorial design over five binary factors: four behavioral predispositions—\emph{loss aversion tendency}, \emph{herding tendency}, \emph{wealth differentiation sensitivity}, and \emph{mispricing sensitivity}—and the \emph{initial trading style} (Technical vs. Fundamental). Each agent is identified by
\[
\theta_i = (\ell_i, h_i, w_i, m_i, \pi_i) \in \{0,1\}^4 \times \{\text{Tech}, \text{Fund}\}.
\]

To operationalize these factors, we prepare persona prompts in either a presence or neutral form. Each prompt specifies a core belief and decision tendencies in trading contexts. Full prompts for all four factors are in Appendix~\ref{prompt:behavioral}.

\noindent\textbf{Initial Wealth and Portfolio Initialization.}  
At the start of the simulation, all agents are endowed with the same initial wealth $W_0$, and each invests half (\(\rho = 0.5\)) in the equity market, with the remaining in cash. Five representative stocks—MSFT, ICE, VRTX, CAT, and CLX—are selected to represent distinct sectors, with each receiving a 10\% allocation of total wealth. This design removes randomness from stock performance and allows behavioral differences to drive outcome variability. Investments are made using pre-simulation split-adjusted closing prices, with positions being long-only and unlevered.

\noindent\textbf{Agent Memory Module.}  
Each agent has a memory module that stores its personality traits, which persist throughout trading and style-switching decisions. It also tracks the agent’s cash balance, stock positions, and entry prices. Two ledgers are initialized: an actual ledger that records real market observations and a counterfactual ledger tracking simulated actions and P\&L(profit and loss) under the alternative style without affecting the portfolio. The memory module is updated daily with actual trades and every 10 trading days with a summary of P\&L, majority-style share, and the agent’s switch-or-stay decision with explanation.

\subsection{Trading Procedure and Decision Process} 
\noindent\textbf{Daily Trading Decision.}  
Each trading day $t$ begins with a decision-making phase using information available up to $t-1$. The agent receives data on all five representative stocks, including price, volume, technical, and fundamental indicators. Based on this and its current holdings, cash, and historical performance, the agent decides to Buy, Sell, or Hold for each stock. A counterfactual decision process is also executed under the opposite trading style to simulate alternative behavior. The decision integrates multiple dimensions, adjusting the portfolio in response to market signals and financial states. Full prompt templates are in Appendix~\ref{prompt:trading}.

\noindent\textbf{Execution and Ledger Recording.}  
Orders are executed at the market open using split-adjusted prices. The system updates the agent's cash balance, positions, and P\&L, recording the transaction in the actual ledger. A counterfactual ledger simulates the agent's performance under the alternate style, with both ledgers synchronized every 10 days for block-level comparisons.

\noindent\textbf{Style Switching Decision.}  
At the end of each 10-day block, the agent evaluates its portfolio performance and market context, considering: (i) current holdings, (ii) available funds, (iii) actual P\&L for the block and year-to-date, (iv) the opposite style's average profit, and (v) the population distribution of styles. Based on this, the agent decides whether to switch to the opposite style, influenced by its personality traits. The decision is accompanied by a natural-language rationale, making it traceable and explainable.

\setcounter{table}{1}
\begin{table*}[t]
  \caption{Technical and fundamental indicators used in the simulation.}
  \label{table-indicators}
  \fontsize{7pt}{10pt}\selectfont
  \setlength{\tabcolsep}{6pt}
  \begin{tabularx}{\textwidth}{l l l X}
    \toprule
    \textbf{Category} & \textbf{Indicator} & \textbf{Formula} & \textbf{Introduction} \\
    \midrule
    \multirow{7}{*}{\textbf{Technical}}
      & close\_price & $C_t$ & Split-adjusted close price on day $t$. \\
      & pre\_close   & $C_{t-1}$ & Prior day's split-adjusted close price. \\
      & change       & $C_t - C_{t-1}$ & Daily price change. \\
      & pct\_chg     & $100 \times \dfrac{C_t - C_{t-1}}{C_{t-1}}$ & Daily percentage change. \\
      & vol          & $\mathrm{Vol}_t$ & Split-adjusted trading volume. \\
      & vol\_N       & $\displaystyle \text{vol}_N(t)=\frac{1}{N}\sum_{k=0}^{N-1}\mathrm{Vol}_{t-k},\; N\in\{5,10,30\}$ & $N$-day average volume (short/medium/long activity). \\
      & ma\_N        & $\displaystyle \text{ma}_N(t)=\frac{1}{N}\sum_{k=0}^{N-1}\mathrm{C}_{t-k},\; N\in\{5,10,30\}$ & Moving average of adjusted close price. \\
    \midrule
    \multirow{4}{*}{\textbf{Fundamental}}
      & Leverage (Debt Ratio) & Liab/Assets & Capital structure / solvency. \\
      & Current Ratio         & CA/CL & Short-term liquidity. \\
      & OCF                   & OCF & Net cash provided by operating activities. \\
      & FCF                   & OCF - CapEx & Discretionary cash flow (approx.). \\
    \bottomrule
  \end{tabularx}
\end{table*}

\section{Experiment}
\subsection{Experimental Setup and Metrics}
\noindent\textbf{Experimental Setup.}  
We simulate the full year of 2024 across 253 trading days on five representative S\&P 500 stocks (MSFT, ICE, VRTX, CAT, and CLX). The agent population consists of \(2^5=32\) heterogeneous agents, constructed via a factorial design based on four behavioral drivers and the initial trading style. We test four different models—GPT-4o-mini, Deepseek-Chat, Gemini-2.5-flash-lite-thinking-8192, and Qwen-2.5-72B-Instruct—as the backbone models for the agents. Trading decisions are made daily, with style reviews every 10 trading days. Implementation details are in the method section.

\noindent\textbf{Evaluation Metrics.}
To assess the impact of each behavioral driver on agents' style-switching behavior, we compute four alignment scores for each driver—loss-aversion, herding, advantage, and mispricing alignment—across two independent cohorts (aligned vs.\ non-aligned). For each driver, we test whether the aligned cohort attains higher scores than the non-aligned cohort using a one-sided Mann--Whitney $U$ test (aligned $>$ non-aligned). We report the test statistic $U$ and its $p$-value, together with three nonparametric effect sizes—rank-biserial correlation, Cliff's $\delta$, and the common-language effect size (CLES)—oriented so that larger values favor the aligned cohort.

For each driver, we compute the aligned-group $U$ statistic from pooled ranks over two independent samples, with $n_A=16$ for aligned agents and $n_B=16$ for non-aligned agents:
\[
U_A=\sum_{j\in A}\mathrm{rank}(x_j)-\frac{n_A(n_A+1)}{2}.
\]

Equivalently, $U_A$ can be interpreted as the number of pairwise wins of aligned over non-aligned agents, with ties counted as $0.5$:
\[
U_A = W + 0.5\,T,
\]
where $W$, $L$, and $T$ denote the numbers of pairwise wins, losses, and ties (aligned vs.\ non-aligned). We obtain the one-sided $p$-value from the large-sample normal approximation to $U$ with tie-variance correction and no continuity correction. Alongside $U$ and $p$-value, we report effect sizes:

\begin{gather*}
\mathrm{CLES} = \frac{W + 0.5\,T}{n_A n_B}, \quad
r_{\mathrm{rb}} = 2 \cdot \mathrm{CLES} - 1, \\
\delta_{\mathrm{Cliff}} = \frac{W - L}{n_A n_B}.
\end{gather*}

At a significance level $\alpha=0.05$, we consider evidence for the hypothesized direction when the one-sided $p<\alpha$; we additionally verify that the effect sizes are consistent with this direction (i.e., $\mathrm{CLES}>0.5$, $r_{\mathrm{rb}}>0$, and $\delta_{\mathrm{Cliff}}>0$).

\begin{table}[h]
  \caption{Mann--Whitney U tests on alignment scores across four models, categorized by behavioral driver. $r\_b$ indicates rank-biserial correlation while $c\_d$ indicates Cliff’s delta.}
  \label{MU_metrics_llms}
  \setlength{\tabcolsep}{1pt}
  \footnotesize
  \begin{tabularx}{\columnwidth}{l>{\centering\arraybackslash}X>{\centering\arraybackslash}X>{\centering\arraybackslash}X>{\centering\arraybackslash}X>{\centering\arraybackslash}X}
    \toprule
    \textbf{Behavioral Drivers} & \textbf{U(\(\uparrow\))} & \textbf{p(\(\downarrow\))} & \textbf{r\_b(\(\uparrow\))} & \textbf{c\_d(\(\uparrow\))} & \textbf{cles(\(\uparrow\))} \\
    \midrule
    \textbf{GPT-4o-mini} \\
    Loss Aversion          & 214.0 & 0.0003 & 0.67 & 0.67 & 0.84 \\
    Herding                & 140.0 & 0.33 & 0.09 & 0.09 & 0.55 \\
    Wealth Differentiation & 174.0 & 0.04 & 0.36 & 0.36 & 0.68 \\
    Price Misalignment     & 155.5 & 0.15 & 0.21 & 0.21 & 0.61 \\
    \midrule
    \textbf{Gemini} \\
    Loss Aversion          & 201.0 & 0.002 & 0.51 & 0.57 & 0.79 \\
    Herding                & 139.0 & 0.34 & 0.09 & 0.09 & 0.54 \\
    Wealth Differentiation & 151.5 & 0.19 & 0.18 & 0.18 & 0.59 \\
    Price Misalignment     & 122.5 & 0.59 & -0.04 & -0.04 & 0.48 \\
    \midrule
    \textbf{DeepSeek} \\
    Loss Aversion          & 195.0 & 0.004 & 0.47 & 0.52 & 0.76 \\
    Herding                & 141.0 & 0.31 & 0.10 & 0.10 & 0.55 \\
    Wealth Differentiation & 131.5 & 0.45 & 0.03 & 0.03 & 0.51 \\
    Price Misalignment     & 119.5 & 0.63 & -0.07 & -0.07 & 0.47 \\
    \midrule
    \textbf{Qwen} \\
    Loss Aversion          & 201.0 & 0.002 & 0.51 & 0.57 & 0.79 \\
    Herding                & 174.0 & 0.04 & 0.36 & 0.36 & 0.68 \\
    Wealth Differentiation & 130.0 & 0.48 & 0.02 & 0.02 & 0.51 \\
    Price Misalignment     & 130.0 & 0.48 & 0.02 & 0.02 & 0.51 \\
    \bottomrule
  \end{tabularx}
\end{table}

\subsection{RQ1 — Loss Aversion Tendency}
\noindent\textbf{Research Question and Intuition.}  
RQ1: Does loss aversion cause an agent to remain in its current style after recent losses? In essence, agents with loss aversion are more likely to avoid switching styles after losses, especially when block-level P\&L falls below key anchors like the break-even point. This yields clear qualitative predictions: more “Stay” decisions following loss blocks, an asymmetric switch threshold around break-even, and a tendency to postpone switching until drawdowns are partially recovered.

\noindent\textbf{Loss-Aversion-Alignment Score (LAS).}
We evaluate agent decisions in 10-day blocks indexed by $b=1,\dots,B$.  
Let \( S_b \in \{\text{Fund}, \text{Tech}\} \) 
denote the strategy adopted at the end of block \( b \). 
We then define the set of switching events as:
\[
\mathrm{Switch}=\{\,b\in\{2,\ldots,B\}: S_{b}\neq S_{b-1}\,\},
\]
which includes all blocks where the strategy changes between consecutive periods.  
Similarly, the set of staying events is defined as:
\[
\mathrm{Stay}=\{\,b\in\{2,\ldots,B\}: S_{b}=S_{b-1}\,\}.
\]

The LAS score is defined for staying events as:
{\footnotesize
$$
\text{LAS}(t) =
\begin{cases}
R_{t-1} - R_t, & \text{if } t \in \mathrm{Stay} \text{ and } R_t < R_{t-1},\\[1ex]
0, & \text{otherwise,}
\end{cases}
$$
}

\noindent where \( R_{t-1} \) and \( R_t \) represent the returns of the \( b_{t-1} \) and \( b_t \) blocks, respectively. Finally, the agent-level LAS is aggregated over all staying events as:
\[
\overline{\text{LAS}} = \frac{1}{|\mathrm{Stay}|} \sum_{t \in \mathrm{Stay}} \text{LAS}(t).
\]

To assess the statistical significance of differences in loss aversion behavior, the 16 agents with strong loss aversion tendency and the 16 agents with weak loss aversion tendency are compared using Mann–Whitney U test.

\noindent\textbf{Findings and theory alignment.}  
Results of the Mann–Whitney \(U\) tests are summarized in Table~\ref{MU_metrics_llms}. All models show significant alignment with loss aversion, but GPT-4o-mini, Gemini, and Qwen display stronger effects, with a higher proportion of aligned agents classified as loss-averse. DeepSeek, however, shows a weaker effect and lower consistency with loss aversion theory, indicating that the model may exhibit less behavioral inertia when faced with losses. These findings support RQ1 and suggest that the switching behavior of LLM agents is influenced by loss aversion in a manner consistent with behavioral finance theory.

\subsection{RQ2 — Herding Tendency}
\noindent\textbf{Research question and intuition.}  
RQ2: Does herding tendency increase switching when the alternative style has a larger population share? Specifically, compared to a herding-neutral agent, does an LLM agent with a herding predisposition switch to the majority style with weaker performance evidence? The mechanism follows social-proof logic: a higher alternative-style share raises the perceived payoff of conformity, lowering the threshold for switching. This is analogous to how individuals in real-world social settings often align with the majority opinion due to perceived benefits of conformity. In our simulator, we broadcast the population counts for each style at the start of each evaluation block, providing group-level information to observe how herding influences agents' switching behavior. We expect that agents with stronger herding tendencies will be more likely to adopt the majority style.

\noindent\textbf{Herd-Alignment Score (HAS)}  
For each switching event $t \in \mathrm{Switch}$, let $n_{\text{other},t}$ be the number of agents adopting the alternative strategy at block $t$, and let $n_{\text{total},t}$ be the total number of agents.  
The Herd-Alignment Score is defined as:
\[
HAS(t) = \frac{n_{\text{other},t}}{n_{\text{total},t}}.
\]

The agent-level HAS is aggregated over all switching events as:
\[
\overline{HAS} = \frac{1}{|\mathrm{Switch}|} \sum_{t \in \mathrm{Switch}} HAS(t).
\]  

Analogous to RQ1, we compare the 16 agents aligned with herding tendency to the 16 non-aligned agents using the Mann--Whitney $U$ test and the main evaluation results of four tested models are reported in Table~\ref{MU_metrics_llms}.

\paragraph{Findings and theory alignment.}  
The results show that Qwen aligns with the hypothesis in RQ2, exhibiting stronger herding tendencies, with a p-value of 0.04, indicating statistical significance. Specifically, herding-aligned Qwen agents tend to switch to the majority style earlier, even with weaker performance evidence, consistent with the social proof behavior observed in real-world scenarios. This suggests that Qwen, when faced with a larger group, is more likely to conform to the majority, aligning with social psychological mechanisms of group behavior. 

In contrast, the other three models—GPT-4o-mini, Gemini, and DeepSeek—do not exhibit significant herding tendencies. For these models, herding-aligned agents do not consistently outperform non-aligned agents in pairwise comparisons, with effect sizes being small and close to random. This suggests that these models fail to incorporate social-driven herd behavior effectively, as agents still prioritize rational decision-making based on evidence rather than conforming to social influences. One possible explanation is that these models focus more on data-driven decisions than on social psychological factors, which hinders their ability to simulate phenomena like market bubbles or herding cascades.

\subsection{RQ3 — Wealth Differentiation Sensitivity}
\noindent\textbf{Research question and intuition.} RQ3: Does wealth differentiation sensitivity trigger switching when the counterfactual style has recently outperformed the current one? In particular, we ask whether sensitivity to relative wealth—the gap between the agent’s realized portfolio value under the current style and the counterfactual ledger it would have earned under the alternative style—induces a catch-up response, causing the agent to be more inclined to switch than an otherwise identical wealth-gap–insensitive agent facing the same signals. 

\noindent\textbf{Advantage-Alignment Score (AAS)}  
For each switching event $t \in \mathrm{Switch}$, 
The absolute advantage difference at block $t$ is:

\[
\Delta R_t = \bar{R}_{t} - R_{t},
\]  
\noindent where $\bar{R}_{t}$ represents the returns of the counterfactual style in block ${b}_{t}$. Then we can define Advantage Alignment Score for this event as:
\[
AAS(t) =
\begin{cases}
\Delta R_t, & \text{if } \Delta R_t > 0,\\[1ex]
0, & \text{otherwise.}
\end{cases}
\]

The agent-level AAS is aggregated as:
\[
\overline{AAS} = \frac{1}{|\mathrm{Switch}|} \sum_{t \in \mathrm{Switch}} AAS(t).
\]

\noindent\textbf{Findings and theory alignment.} 
The results show that GPT-4o-mini aligns with the hypothesis in RQ3, exhibiting a strong and consistent response to wealth differentiation. Specifically, wealth-differentiation-aligned GPT-4o-mini agents consistently show higher wealth alignment compared to non-aligned agents, with a moderate effect size and a clear advantage in pairwise comparisons. This suggests that GPT-4o-mini can internalize relative wealth comparisons as a stable heuristic for switching, supporting the idea that agents with a sensitivity to wealth differentiation are more likely to align with wealth-based performance signals.

However, the other models—Gemini, DeepSeek, and Qwen—do not exhibit consistent wealth differentiation alignment. For these models, wealth-differentiation-aligned agents do not reliably outperform non-aligned agents, and the effect sizes are small. This suggests that these models fail to incorporate wealth differentiation as a stable driving force for switching behavior. One possible explanation is that, unlike GPT-4o-mini, these models may prioritize other performance factors or rely on more generic decision-making heuristics that do not emphasize relative wealth gaps. As a result, wealth-driven style migration does not emerge strongly in these models, limiting their ability to replicate the wealth-based switching behavior observed in GPT-4o-mini.

\subsection{RQ4 — Price Misalignment Sensitivity}
\noindent\textbf{Research question and intuition.} RQ4: does price misalignment sensitivity induce shifts toward the fundamental style when market price diverges from estimated fundamental value? In particular, we ask whether sensitivity to the price–value gap makes an LLM agent tilt toward the fundamental style sooner than an otherwise identical price-gap–insensitive agent facing the same signals. The intuition is mean-reversion: as the absolute misalignment grows, the expected payoff from fundamentals-driven bets rises relative to trend-following cues, lowering the internal threshold for abandoning the current style.

\noindent\textbf{Mispricing-Alignment Score (MAS).} 
At each switching block $t$, we first select another stock from the same sector as an auxiliary reference and estimate a sector-level time-series regression that links contemporaneous valuation (P/E proxied by market cap/OCF) to the realized 20-day forward return. This provides an interpretable price--value mapping. We then apply the fitted model to the five-stock universe described in the experimental setup to obtain predicted returns $\hat{Y}$, and define the short-horizon mispricing magnitude as the absolute prediction error:
\[
Y = \alpha + \beta X, \quad MAS(t) = |\hat{Y} - Y|,
\]
where $X$ is the P/E proxy and $Y$ is the realized 20-day forward return. To reflect switching direction, we aggregate MAS over switching events as:
\[
\overline{MAS} =
\sum_{\substack{t \in \mathrm{Switch} \\ \text{Tech} \to \text{Fund}}} MAS(t)
-
\sum_{\substack{t \in \mathrm{Switch} \\ \text{Fund} \to \text{Tech}}} MAS(t).
\]

\noindent\textbf{Findings and theory alignment.}
Results show that price misalignment sensitivity does not yield robust mispricing alignment across models. For all models, mispricing-aligned agents do not consistently achieve higher score than non-aligned agents. This suggests that, LLM agents do not reliably exhibit the migration predicted by behavioral finance, namely switching toward fundamentals as the price--value gap widens. One possible explanation is that valuation-based signals are outweighed by more salient local cues, weakening systematic mispricing-driven switching.

\subsection{Robustness Evaluation}
To assess the robustness of our evaluation framework, we conducted two additional experiments. We first used a rule-based ABM model inspired by \citet{traditional-ABM3} to verify whether the rules based on traditional financial literature produce significant results within our framework. The results in Table~\ref{MU_metrics_backbones} show that all four behavioral drivers align with theoretical expectations in the traditional ABM simulation. This suggests that, despite its simplicity and lower complexity, the rule-based ABM still produces results consistent with traditional financial theory. This also validates the robustness of our framework, since the traditional ABM framework, based on classical financial theory, does indeed show significant results.

Next, we conducted an ablation experiment to test whether our consistency verification framework still produces significant results without psychological factors. In this experiment, agents were randomly grouped with no psychological drivers, which reduces the system to a basic LLM-ABM model. As expected, since random grouping should not result in systematic differences, no significant effects were found for the four behavioral drivers. This result further validates the robustness of our consistency verification framework.

\begin{table}[t!]
  \caption{Mann–Whitney U tests on alignment scores by behavioral driver in experiments with the traditional ABM and the LLM-ABM without behavioral factors. For simplicity, we only report the results from GPT-4o-mini for LLM-ABM-w\_drivers. $r\_b$ indicates rank-biserial correlation while $c\_d$ indicates Cliff’s delta.}
  \label{MU_metrics_backbones}
  \setlength{\tabcolsep}{1pt}
  \footnotesize
  \begin{tabularx}{\columnwidth}{l>{\centering\arraybackslash}X>{\centering\arraybackslash}X>{\centering\arraybackslash}X>{\centering\arraybackslash}X>{\centering\arraybackslash}X}
    \toprule
    \textbf{Behavioral Drivers} & \textbf{U(\(\uparrow\))} & \textbf{p(\(\downarrow\))} & \textbf{r\_b(\(\uparrow\))} & \textbf{c\_d(\(\uparrow\))} & \textbf{cles(\(\uparrow\))} \\
    \midrule
    \textbf{LLM-ABM-w\_drivers} \\
    Loss Aversion          & 214.0 & 0.0003 & 0.67 & 0.67 & 0.84 \\
    Herding                & 140.0 & 0.33 & 0.09 & 0.09 & 0.55 \\
    Wealth Differentiation & 174.0 & 0.04 & 0.36 & 0.36 & 0.68 \\
    Price Misalignment     & 155.5 & 0.15 & 0.21 & 0.21 & 0.61 \\
    \midrule
    \textbf{ABM} \\
    Loss Aversion          & 200.0 & 0.003 & 0.48 & 0.56 & 0.78 \\
    Herding                & 206.5 & 0.002 & 0.52 & 0.61 & 0.81 \\
    Wealth Differentiation & 222.0 & 0.0003 & 0.72 & 0.85 & 0.93 \\
    Price Misalignment     & 185.0 & 0.016 & 0.38 & 0.45 & 0.72 \\
    \midrule
    \textbf{LLM-ABM-w/o\_drivers} \\
    Loss Aversion & 134.0 & 0.41 & 0.04 & 0.05 & 0.52 \\
    Herding & 138.0 & 0.35 & 0.08 & 0.08 & 0.54 \\
    Wealth Differentiation & 118.5 & 0.65 & -0.07 & -0.07 & 0.46 \\
    Price Misalignment & 125.5 & 0.55 & -0.02 & -0.02 & 0.49 \\
    \bottomrule
  \end{tabularx}
\end{table}

\section{Conclusion}
This study provides a framework for evaluating the behavioral consistency of LLM agents in financial market simulations, focusing on their ability to switch strategies in alignment with established financial theories. In year-long simulations, agents process daily price-volume data, trade under a designated style, and reassess their strategy every 10 trading days. By operationalizing key financial behavioral drivers such as loss aversion, herding, wealth differentiation, and price misalignment, we show that LLM agents' behavior aligns with real market dynamics to some extent, although not consistently across all drivers. Their decision-making remains primarily driven by rational, short-term considerations, particularly in the case of herding and price misalignment. This highlights the need to explore better methods for embedding personality traits, in order to more accurately capture real-world market behavior. Future work can build on this framework to further refine agent behavior, enabling more accurate and dynamic simulations of financial markets.

\section*{Limitations}
One key limitation of this study lies in the absence of an order-matching mechanism in our framework. The primary focus of this paper is to validate the behavioral consistency of LLM agents, specifically examining how well their style-switching behavior aligns with established financial theory. To this end, we do not model the interaction between agents that would traditionally drive price formation through order matching. Instead, we simplify the simulation by directly providing the real price from the previous trading day as the current day’s opening price. This approach is intentional, as our goal is not to simulate the full market dynamics, including phenomena like random volatility, price formation, or market bubbles, but rather to focus on behavioral consistency and alignment with financial theory.

While this design choice helps isolate the specific impact of behavioral drivers on agents' decision-making, it also means that our simulations do not reproduce macro-level phenomena typically observed in financial markets, such as endogenous price formation or the emergent behaviors driven by the collective decisions of agents. Consequently, comparisons between our framework and more complex models, which include such dynamics, should be made cautiously, as they are not directly comparable in terms of their ability to replicate broader market phenomena.

Future work could explore the inclusion of order-matching mechanisms, allowing agents to influence price formation and test whether this enables the simulation of more realistic market dynamics, including the replication of phenomena like market bubbles or herd-driven price movements.



\bibliography{sample}

@article{schelling,
  title={Dynamic models of segregation},
  author={Schelling, Thomas C},
  journal={Journal of mathematical sociology},
  volume={1},
  number={2},
  pages={143--186},
  year={1971},
  publisher={Taylor \& Francis}
}

@book{epstein,
  title={Growing artificial societies: social science from the bottom up},
  author={Epstein, Joshua M and Axtell, Robert},
  year={1996},
  publisher={Brookings Institution Press}
}

@article{bonabeau,
  title={Agent-based modeling: Methods and techniques for simulating human systems},
  author={Bonabeau, Eric},
  journal={Proceedings of the national academy of sciences},
  volume={99},
  number={suppl\_3},
  pages={7280--7287},
  year={2002},
  publisher={National Academy of Sciences}
}

@article{lu2024llms,
  title={Llms and generative agent-based models for complex systems research},
  author={Lu, Yikang and Aleta, Alberto and Du, Chunpeng and Shi, Lei and Moreno, Yamir},
  journal={Physics of Life Reviews},
  volume={51},
  pages={283--293},
  year={2024},
  publisher={Elsevier}
}

@article{zhao2023survey,
  title={A survey of large language models},
  author={Zhao, Wayne Xin and Zhou, Kun and Li, Junyi and Tang, Tianyi and Wang, Xiaolei and Hou, Yupeng and Min, Yingqian and Zhang, Beichen and Zhang, Junjie and Dong, Zican and others},
  journal={arXiv preprint arXiv:2303.18223},
  volume={1},
  number={2},
  year={2023}
}

@article{shao2023character,
  title={Character-llm: A trainable agent for role-playing},
  author={Shao, Yunfan and Li, Linyang and Dai, Junqi and Qiu, Xipeng},
  journal={arXiv preprint arXiv:2310.10158},
  year={2023}
}

@article{li2023camel,
  title={Camel: Communicative agents for" mind" exploration of large language model society},
  author={Li, Guohao and Hammoud, Hasan and Itani, Hani and Khizbullin, Dmitrii and Ghanem, Bernard},
  journal={Advances in Neural Information Processing Systems},
  volume={36},
  pages={51991--52008},
  year={2023}
}

@inproceedings{park2023generative,
  title={Generative agents: Interactive simulacra of human behavior},
  author={Park, Joon Sung and O'Brien, Joseph and Cai, Carrie Jun and Morris, Meredith Ringel and Liang, Percy and Bernstein, Michael S},
  booktitle={Proceedings of the 36th annual acm symposium on user interface software and technology},
  pages={1--22},
  year={2023}
}

@inproceedings{kovavrik2025game,
  title={Game Theory with Simulation in the Presence of Unpredictable Randomisation},
  author={Kova{\v{r}}{\'\i}k, Vojt{\v{e}}ch and Sauerberg, Nathaniel and Hammond, Lewis and Conitzer, Vincent},
  booktitle={Proceedings of the 24th International Conference on Autonomous Agents and Multiagent Systems},
  pages={1191--1199},
  year={2025}
}

@inproceedings{zhang2024multimodal,
  title={A multimodal foundation agent for financial trading: Tool-augmented, diversified, and generalist},
  author={Zhang, Wentao and Zhao, Lingxuan and Xia, Haochong and Sun, Shuo and Sun, Jiaze and Qin, Molei and Li, Xinyi and Zhao, Yuqing and Zhao, Yilei and Cai, Xinyu and others},
  booktitle={Proceedings of the 30th acm sigkdd conference on knowledge discovery and data mining},
  pages={4314--4325},
  year={2024}
}

@article{wu2023bloomberggpt,
  title={Bloomberggpt: A large language model for finance},
  author={Wu, Shijie and Irsoy, Ozan and Lu, Steven and Dabravolski, Vadim and Dredze, Mark and Gehrmann, Sebastian and Kambadur, Prabhanjan and Rosenberg, David and Mann, Gideon},
  journal={arXiv preprint arXiv:2303.17564},
  year={2023}
}

@article{lai2024sec,
  title={Sec-qa: A systematic evaluation corpus for financial qa},
  author={Lai, Viet Dac and Krumdick, Michael and Lovering, Charles and Reddy, Varshini and Schmidt, Craig and Tanner, Chris},
  journal={arXiv preprint arXiv:2406.14394},
  year={2024}
}

@article{xiao2024tradingagents,
  title={TradingAgents: Multi-agents LLM financial trading framework},
  author={Xiao, Yijia and Sun, Edward and Luo, Di and Wang, Wei},
  journal={arXiv preprint arXiv:2412.20138},
  year={2024}
}

@article{ABM-survey1,
  title={Agent-based modeling in economics and finance: Past, present, and future},
  author={Axtell, Robert L and Farmer, J Doyne},
  journal={Journal of Economic Literature},
  volume={63},
  number={1},
  pages={197--287},
  year={2025},
  publisher={American Economic Association 2014 Broadway, Suite 305, Nashville, TN 37203-2425}
}

@article{traditional-ABM1,
  title={Artificial economic life: a simple model of a stockmarket},
  author={Palmer, Richard G and Arthur, W Brian and Holland, John H and LeBaron, Blake and Tayler, Paul},
  journal={Physica D: Nonlinear Phenomena},
  volume={75},
  number={1-3},
  pages={264--274},
  year={1994},
  publisher={Elsevier}
}

@incollection{traditional-ABM2,
  title={Asset pricing under endogenous expectations in an artificial stock market},
  author={Arthur, W Brian and Holland, John H and LeBaron, Blake and Palmer, Richard and Tayler, Paul},
  booktitle={The economy as an evolving complex system II},
  pages={15--44},
  year={2018},
  publisher={CRC Press}
}

@article{traditional-ABM3,
  title={Scaling and criticality in a stochastic multi-agent model of a financial market},
  author={Lux, Thomas and Marchesi, Michele},
  journal={Nature},
  volume={397},
  number={6719},
  pages={498--500},
  year={1999},
  publisher={Nature Publishing Group UK London}
}

@article{traditional-ABM4,
  title={The price dynamics of common trading strategies},
  author={Farmer, J Doyne and Joshi, Shareen},
  journal={Journal of Economic Behavior \& Organization},
  volume={49},
  number={2},
  pages={149--171},
  year={2002},
  publisher={Elsevier}
}

@article{traditional-ABM5,
  title={A microsimulation of traders activity in the stock market: the role of heterogeneity, agents’ interactions and trade frictions},
  author={Iori, Giulia},
  journal={Journal of Economic Behavior \& Organization},
  volume={49},
  number={2},
  pages={269--285},
  year={2002},
  publisher={Elsevier}
}

@article{LLM-ABM-fin-eco-simulation1,
  title={Econagent: large language model-empowered agents for simulating macroeconomic activities},
  author={Li, Nian and Gao, Chen and Li, Mingyu and Li, Yong and Liao, Qingmin},
  journal={arXiv preprint arXiv:2310.10436},
  year={2023}
}

@article{LLM-ABM-fin-eco-simulation2,
  title={Simulating Macroeconomic Expectations using LLM Agents},
  author={Lin, Jianhao and Sun, Lexuan and Yan, Yixin},
  journal={arXiv preprint arXiv:2505.17648},
  year={2025}
}

@article{LLM-ABM-survey1,
  title={Large language models empowered agent-based modeling and simulation: A survey and perspectives},
  author={Gao, Chen and Lan, Xiaochong and Li, Nian and Yuan, Yuan and Ding, Jingtao and Zhou, Zhilun and Xu, Fengli and Li, Yong},
  journal={Humanities and Social Sciences Communications},
  volume={11},
  number={1},
  pages={1--24},
  year={2024},
  publisher={Palgrave}
}

@inproceedings{LLM-ABM-survey2,
  title={Large Language Model Based Multi-agents: A Survey of Progress and Challenges},
  author={Guo, Taicheng and Chen, Xiuying and Wang, Yaqi and Chang, Ruidi and Pei, Shichao and Chawla, Nitesh V and Wiest, Olaf and Zhang, Xiangliang},
  booktitle={IJCAI},
  year={2024}
}

@article{LLM-ABM-stock-market1-twinmarket,
  title={TwinMarket: A Scalable Behavioral and Social Simulation for Financial Markets},
  author={Yang, Yuzhe and Zhang, Yifei and Wu, Minghao and Zhang, Kaidi and Zhang, Yunmiao and Yu, Honghai and Hu, Yan and Wang, Benyou},
  journal={arXiv preprint arXiv:2502.01506},
  year={2025}
}

@article{LLM-ABM-stock-market2-cantrade,
  title={Can Large Language Models Trade? Testing Financial Theories with LLM Agents in Market Simulations},
  author={Lopez-Lira, Alejandro},
  journal={arXiv preprint arXiv:2504.10789},
  year={2025}
}

@article{LLM-ABM-stock-market3-stockagent,
  title={When AI Meets Finance (StockAgent): Large Language Model-based Stock Trading in Simulated Real-world Environments},
  author={Zhang, Chong and Liu, Xinyi and Jin, Mingyu and Zhang, Zhongmou and Li, Lingyao and Wang, Zhenting and Hua, Wenyue and Shu, Dong and Zhu, Suiyuan and Jin, Xiaobo and others},
  journal={CoRR},
  year={2024}
}

@article{LLM-ABM-stock-market4-gao,
  title={Simulating Financial Market via Large Language Model based Agents},
  author={Gao, Shen and Wen, Yuntao and Zhu, Minghang and Wei, Jianing and Cheng, Yuhan and Zhang, Qunzi and Shang, Shuo},
  journal={CoRR},
  year={2024}
}

@article{LLM-ABM-stock-market5-fincon,
  title={Fincon: A synthesized llm multi-agent system with conceptual verbal reinforcement for enhanced financial decision making},
  author={Yu, Yangyang and Yao, Zhiyuan and Li, Haohang and Deng, Zhiyang and Jiang, Yuechen and Cao, Yupeng and Chen, Zhi and Suchow, Jordan and Cui, Zhenyu and Liu, Rong and others},
  journal={Advances in Neural Information Processing Systems},
  volume={37},
  pages={137010--137045},
  year={2024}
}

@inproceedings{LLM-ABM-stock-market6-shiftingpower,
  title={Shifting Power: Leveraging LLMs to Simulate Human Aversion in ABMs of Bilateral Financial Exchanges, A bond market study},
  author={Vidler, Alicia and Walsh, Toby},
  booktitle={Proceedings of the 24th International Conference on Autonomous Agents and Multiagent Systems},
  pages={2777--2779},
  year={2025}
}

@article{LLM-ABM-stock-market7-henning,
  title={LLM Trading: Analysis of LLM Agent Behavior in Experimental Asset Markets},
  author={Henning, Thomas and Ojha, Siddhartha M and Spoon, Ross and Han, Jiatong and Camerer, Colin F},
  journal={arXiv preprint arXiv:2502.15800},
  year={2025}
}

@article{behavioral-drivers1-Brock,
  title={A rational route to randomness},
  author={Brock, William A and Hommes, Cars H},
  journal={Econometrica: Journal of the Econometric Society},
  pages={1059--1095},
  year={1997},
  publisher={JSTOR}
}

@article{behavioral-drivers2-Brock,
  title={Heterogeneous beliefs and routes to chaos in a simple asset pricing model},
  author={Brock, William A and Hommes, Cars H},
  journal={Journal of Economic dynamics and Control},
  volume={22},
  number={8-9},
  pages={1235--1274},
  year={1998},
  publisher={Elsevier}
}

@article{behavioral-drivers3-structural,
  title={Structural stochastic volatility in asset pricing dynamics: Estimation and model contest},
  author={Franke, Reiner and Westerhoff, Frank},
  journal={Journal of Economic Dynamics and Control},
  volume={36},
  number={8},
  pages={1193--1211},
  year={2012},
  publisher={Elsevier}
}

@article{behavioral-drivers4-lossaversion,
  title={Prospect theory: An analysis of decision under risk},
  author={Kai-Ineman, DANIEL and Tversky, Amos and others},
  journal={Econometrica},
  volume={47},
  number={2},
  pages={363--391},
  year={1979}
}

@article{behavioral-drivers5-herding,
  title={A simple model of herd behavior},
  author={Banerjee, Abhijit V},
  journal={The quarterly journal of economics},
  volume={107},
  number={3},
  pages={797--817},
  year={1992},
  publisher={MIT Press}
}

@article{behavioral-drivers6-differential,
  title={Heterogeneous agent models in economics and finance},
  author={Hommes, Cars H},
  journal={Handbook of computational economics},
  volume={2},
  pages={1109--1186},
  year={2006},
  publisher={Elsevier}
}

@article{behavioral-drivers7-misalignment,
  title={Stock prices, earnings, and expected dividends},
  author={Campbell, John Y and Shiller, Robert J},
  journal={the Journal of Finance},
  volume={43},
  number={3},
  pages={661--676},
  year={1988},
  publisher={Wiley Online Library}
}

@techreport{IT,
  author       = {{S\&P Dow Jones Indices LLC}},
  title        = {Global Sector Primer Series: Information Technology},
  institution  = {S\&P Dow Jones Indices LLC},
  year         = {2022},
  address      = {New York},
  url          = {https://www.spglobal.com/spdji/en/documents/education/education-global-sector-primer-series-information-technology.pdf}
}

@article{Finance,
  author       = {Kashyap, Anil K. and Stein, Jeremy C.},
  title        = {What Do a Million Observations on Banks Say about the Transmission of Monetary Policy?},
  journal      = {American Economic Review},
  year         = {2000},
  volume       = {90},
  number       = {3},
  pages        = {407--428},
  doi          = {10.1257/aer.90.3.407},
  url          = {https://www.aeaweb.org/articles?id=10.1257/aer.90.3.407}
}

@techreport{Health,
  author       = {{S\&P Dow Jones Indices LLC}},
  title        = {Global Sector Primer Series: Health Care},
  institution  = {S\&P Dow Jones Indices LLC},
  year         = {2021},
  address      = {New York},
  url          = {https://www.spglobal.com/spdji/en/documents/education/education-global-sector-primer-series-health-care.pdf}
}

@techreport{IndustrialConsumer,
  author       = {{MSCI Inc.}},
  title        = {MSCI Cyclical and Defensive Sectors Indexes Methodology},
  institution  = {MSCI Inc.},
  year         = {2022},
  month        = {November},
  address      = {New York},
  url          = {https://www.msci.com/eqb/methodology/meth_docs/MSCI_Cyclical_and_Defensive_Sectors_Indexes_Methodology_Nov22.pdf}
}

\appendix

\setcounter{table}{0}
\begin{table*}[t!]
  \caption{Data fields used to compute technical and fundamental indicators}
  \label{table-data-fields}
  \fontsize{7pt}{10pt}\selectfont
  \setlength{\tabcolsep}{6pt}
  \begin{tabularx}{\textwidth}{l l l X}
    \toprule
    \textbf{Data Source Category} & \textbf{Field} & \textbf{Abbr.} & \textbf{Introduction} \\
    \midrule
    \multirow{5}{*}{\textbf{Daily Stock Data}}
      & Date                                           & Date  & U.S. trading day (America/New\_York). \\
      & Open, High, Low, Close (split-adjusted)        & OHLC  & Daily open, high, low, and close prices (split-adjusted). \\
      & Volume (split-adjusted)                        & Vol   & Number of shares traded (adjusted for splits). \\
      & Dividends                                      & Div   & Cash dividends per share. \\
      & Stock Splits                                   & Splits& Split ratio on the date; used to adjust OHLC and Volume. \\
    \midrule
    \multirow{6}{*}{\textbf{Corporate Disclosures}}
      & Assets                                         & Assets& Total assets; snapshot of firm resources. \\
      & Liabilities                                    & Liab  & Total liabilities; obligations to creditors. \\
      & AssetsCurrent                                  & CA    & Current assets; expected to be realized within one year. \\
      & LiabilitiesCurrent                             & CL    & Current liabilities; obligations due within one year. \\
      & NetCashProvidedByUsedInOperatingActivities     & OCF   & Operating cash flow; cash generated by core operations. \\
      & PaymentsToAcquirePropertyPlantAndEquipment     & CapEx & Capital expenditures for PP\&E; long-term outflows. \\
    \bottomrule
  \end{tabularx}
\end{table*}

\section{Details for Simulation Setup.}
\label{simulation_setup}
\subsection{Implementation Details}
\label{implementation_details}
To avoid pre-exposure bias, all features for trading day $t$ are computed from information timestamped $\le t{-}1$; prices and volumes are split-adjusted, and fundamentals follow a disclosure-lag policy (usable only after the filing date). Rolling indicators are features computed over trailing windows of length $N$ (e.g., $N$-day moving/averaged quantities). Because the simulation starts on the first trading day in 2024, we adopt a warm-up equal to the longest window (30 trading days): before $N$ prior observations have accumulated, such features are set to \textsc{NA} and are not exposed to the agent.

\subsection{Stock Pool Construction Details}
\label{stock_pool_details}
Starting from the S\&P 500, we first screen firms for disclosure completeness and then, within the GICS taxonomy, select five sectors—Information Technology, Financials, Health Care, Industrials, and Consumer Staples—and choose one high-quality representative from each to form the simulation pool: MSFT, ICE, VRTX, CAT, and CLX. These sectors are functionally complementary: Information Technology reflects innovation-driven, higher-beta growth~\cite{IT}; Financials transmit liquidity conditions and provide market infrastructure~\cite{Finance}; Health Care combines defensiveness with dense R\&D/regulatory events~\cite{Health}; Industrials are tied to capex and logistics cycles~\cite{IndustrialConsumer}; and Staples offer stable cash flows and inelastic demand~\cite{IndustrialConsumer}. This composition spans both defensive and cyclical/growth exposures and yields a heterogeneous information set grounded in corporate disclosures and price–volume data, thereby providing agents with sufficient variation to trade.

\subsection{Details for Data Fields}
We collected data for the fields shown in Table~\ref{table-data-fields}, which can be divided into two categories: Daily Stock Data and Corporate Disclosure.

The \textbf{Daily Stock Data} category includes several key data fields essential for analyzing stock performance. \textbf{Date} field represents the trading day. The \textbf{Open, High, Low, Close (OHLC)} prices reflect the daily price movements, with adjustments made for stock splits to ensure consistency. The \textbf{Volume (Vol)} field records the number of shares traded during the day, with adjustments for splits to account for changes in the stock's share count. \textbf{Dividends (Div)} represent the cash dividends paid per share, providing insights into the income return for investors. \textbf{Stock Splits (Splits)} indicate the split ratio on the given date and are used to adjust the OHLC prices and volume data, ensuring accurate historical comparisons.

The \textbf{Corporate Disclosures} category includes financial data provided by companies to give insights into their financial position. \textbf{Assets} represent the total assets of a company, providing a snapshot of its resources. \textbf{Liabilities (Liab)} reflect the total obligations the company has to its creditors. \textbf{AssetsCurrent (CA)} includes assets expected to be realized within one year, offering a view of short-term financial health. Similarly, \textbf{LiabilitiesCurrent (CL)} represents obligations due within one year, indicating the company’s short-term liabilities. \textbf{NetCashProvidedByUsedInOperatingActivities (OCF)} represents the cash flow generated by the company’s core operations, which is critical for assessing operational efficiency. Lastly, \textbf{PaymentsToAcquirePropertyPlantAndEquipment (CapEx)} refers to capital expenditures for property, plant, and equipment, indicating the company’s long-term investment in its physical assets.

\section{Prompts}\label{appendix-A}
\subsection{Prompts for Behavioral Predispositions}\label{prompt:behavioral}

\noindent\textbf{Loss Aversion Tendency.}
Loss aversion is a behavioral bias whereby individuals perceive losses more intensely than equal-sized gains. This bias affects decision making, often prompting premature profit-taking, reluctance to cut losses, and heightened sensitivity to reference points such as the purchase price or recent highs/lows. In the context of style switching, a loss-averse trader tends to persist with the current style after drawdowns, delaying or avoiding switches even when the opposite style has sustained a counterfactual advantage over recent evaluation blocks. The figure below presents exemplar prompts corresponding to strong/consistent and weak/inconsistent variants of this predisposition; these prompts encode a core belief and concrete switching tendencies.

\Needspace{18\baselineskip}

\begin{tcolorbox}[title=Loss Aversion (Strong / Consistent), fontupper=\footnotesize, breakable]
\textbf{Core belief:} Losses of the same magnitude feel more painful than equal gains; outcomes are assessed against reference points such as purchase price and recent highs/lows.

\vspace{0.5mm}
\textbf{Behavioral tendencies:}
\begin{itemize}[leftmargin=*, itemsep=0.2ex, parsep=0pt, topsep=0.2ex, partopsep=0pt]
  \item Realize profits early to avoid ``giving back'' gains.
  \item Hold or even add to losers, hoping to ``get back to even.''
  \item Strong reluctance to convert paper losses into realized losses.
  \item After losses or sharp volatility, postpone major switches and decisions.
  \item Overweight break-even/anchor prices (``get-out price,'' ``cost basis'').
  \item Prefer options that minimize short-term psychological pain even if long-run expectancy is inferior.
\end{itemize}
\end{tcolorbox}

\begin{tcolorbox}[title=Loss Aversion (Weak / Inconsistent), fontupper=\footnotesize, breakable]
\textbf{Core belief:} Gains and losses carry roughly symmetric psychological weights; decisions are led by expected return/risk and verifiable evidence.

\vspace{0.5mm}
\textbf{Behavioral tendencies:}
\begin{itemize}[leftmargin=*, itemsep=0.2ex, parsep=0pt, topsep=0.2ex, partopsep=0pt]
  \item Let winners run within risk limits instead of exiting prematurely.
  \item Cut losers decisively when evidence indicates a mistake; do not anchor on entry price.
  \item Treat sunk costs and reference prices as irrelevant variables.
  \item Tolerate short-term drawdowns to pursue long-term statistical edge.
  \item Switch strategies based on advantage consistency, not emotions.
  \item Evaluate at the portfolio level to reduce single-trade noise.
\end{itemize}
\end{tcolorbox}

\captionof{figure}{Prompts — Loss Aversion Tendency}
\par\medskip

\noindent\textbf{Herding Tendency.}
Herding behavior refers to the systematic tendency to align one’s decisions with the majority, placing social consensus above independent assessment. In financial markets, this pushes traders to emulate others’ trades, amplifying trends and increasing the likelihood of bubbles and self-reinforcing boom–bust cycles. Mechanistically, herding raises the weight placed on social signals (e.g., perceived majority positions) relative to private information, increasing the chance of information cascades when early movers align. In the context of style switching, herding-prone traders are more likely to consult the population share of styles and migrate toward the prevailing style. The figure below presents exemplar prompts designed to elicit herding behavior in an LLM agent; the two panels correspond to strong/consistent and weak/inconsistent expressions of this predisposition.


\begin{tcolorbox}[title=Herding (Strong / Consistent), 
before=\vspace{0.2\baselineskip},
fontupper=\footnotesize, breakable]
\textbf{Core belief:} Majority behavior conveys information and a sense of safety; aligning with the crowd reduces reputational and regret risks of being ``wrong alone.''

\vspace{0.5mm}
\textbf{Behavioral tendencies:}
\begin{itemize}[leftmargin=*, itemsep=0.2ex, parsep=0pt, topsep=0.2ex, partopsep=0pt]
  \item Use ``social proof/consensus'' as a key signal under uncertainty, down-weighting private information.
  \item Follow when broad agreement or a dominant narrative appears, even if personal evidence is only moderate.
  \item Avoid minority positions to reduce potential loss and psychological pressure.
  \item Calibrate priors with peer views; trust ``hot stories'' and mainstream frames more easily.
  \item Seek emotional safety from conformity and ``post-hoc justification.''
  \item Focus on ``what others do / how the market sees it,'' weakening audits of one’s own model.
\end{itemize}
\end{tcolorbox}

\begin{tcolorbox}[title=Herding (Weak / Inconsistent), fontupper=\footnotesize, breakable]
\textbf{Core belief:} Independent information and one’s own model take priority; majority views may contain synchronized bias and amplified noise.

\vspace{0.5mm}
\textbf{Behavioral tendencies:}
\begin{itemize}[leftmargin=*, itemsep=0.2ex, parsep=0pt, topsep=0.2ex, partopsep=0pt]
  \item Rely on evidence and reasoning first; treat crowd views only as background context.
  \item Willingly take minority positions when evidence is sufficient; withstand asynchrony with the crowd.
  \item Audit fashionable narratives for falsifiability; beware ``information cascades'' and stampedes.
  \item Decouple consensus from decision quality; do not trade consistency for emotional safety.
  \item Curb FOMO using causal reasoning and data consistency.
  \item Maintain metacognitive checks on model and evidence to avoid being swayed by sentiment intensity.
\end{itemize}
\end{tcolorbox}

\captionof{figure}{Prompts — Herding Tendency}
\par\medskip

\noindent\textbf{Wealth Differentiation Sensitivity.}
Wealth differentiation sensitivity refers to the tendency to adjust decisions based on relative performance—comparing one’s realized or prospective returns to an alternative strategy. Mechanistically, the agent treats the realized gap between its current style and the counterfactual style as evidence of misallocation; larger and more persistent gaps raise the posterior in favor of switching. In the context of style switching, such traders are more likely to abandon the current style when the opposite style accrues a positive counterfactual advantage. The figure below presents exemplar prompts eliciting this predisposition; the two panels correspond to strong/consistent and weak/inconsistent variants.

\Needspace{18\baselineskip}

\begin{tcolorbox}[title=Wealth Differentiation Sensitivity (Strong / Consistent), before=\vspace{0.2\baselineskip},fontupper=\footnotesize, breakable]
\textbf{Core belief:} ``How wealthy I would be on the alternative path/style'' is a major reference; position relative to that imagined path shapes safety and self-evaluation.

\vspace{0.5mm}
\textbf{Behavioral tendencies:}
\begin{itemize}[leftmargin=*, itemsep=0.2ex, parsep=0pt, topsep=0.2ex, partopsep=0pt]
  \item Highly sensitive to ``I’d be richer/poorer if I had taken the other route,'' prone to opportunity-loss regret.
  \item When feeling ``behind'' the counterfactual path, favor catch-up choices; when ``ahead,'' prefer continuation.
  \item Read accumulated past outcomes as proof/refutation of personal judgment; style switches feel like endorsing or overturning the past.
  \item Frequently monitor the gap to the imagined path; shrinking that gap becomes a hidden goal even without present statistical support.
  \item Define success by ``making up the past gap,'' down-weighting current risks and evidence quality.
\end{itemize}
\end{tcolorbox}

\vspace{-10pt}

\begin{tcolorbox}[title=Wealth Differentiation Sensitivity (Weak / Inconsistent), fontupper=\footnotesize, breakable]
\textbf{Core belief:} Wealth on an alternative path is background only; current choices are guided by forward-looking risk--return, not counterfactual anchors.

\vspace{0.5mm}
\textbf{Behavioral tendencies:}
\begin{itemize}[leftmargin=*, itemsep=0.2ex, parsep=0pt, topsep=0.2ex, partopsep=0pt]
  \item Avoid hindsight and counterfactual regret as decision drivers.
  \item Separate past path outcomes from current selection; do not aim to ``prove the past right'' or ``recoup past errors.''
  \item Keep emotional distance from the ``gap to the imagined path''; focus on verifiable edge, consistency, and constraints now.
  \item Willing to diverge from prior practice or popular narratives when evidence warrants.
  \item Define success by portfolio-level long-run expectation and risk control, not by closing historical gaps.
\end{itemize}
\end{tcolorbox}

\captionof{figure}{Prompts — Wealth Differentiation Sensitivity}
\par\medskip

\noindent\textbf{Price Misalignment Sensitivity.}
Price misalignment sensitivity is the tendency to adjust decisions when market prices diverge from an estimate of fundamental value, treating large or persistent gaps as evidence of mispricing. Mechanistically, the agent tracks a price--value gap (and its recent change) derived from parsimonious fundamentals and interprets widening gaps as stronger corrective evidence. In the context of style switching, misalignment-sensitive traders are more likely to rotate into the Fundamental style as the gap widens, and to favor Technical as it narrows or reverses. The figure below presents exemplar prompts for this predisposition; the two panels correspond to strong/consistent and weak/inconsistent variants.

\Needspace{18\baselineskip}

\begin{tcolorbox}[title=Price Misalignment Sensitivity (Strong / Consistent), fontupper=\footnotesize, breakable]
\textbf{Core belief:} Large deviations from fundamentals are unsustainable; the larger the gap, the stronger the pull to revert toward fundamental anchors.

\vspace{0.5mm}
\textbf{Behavioral tendencies:}
\begin{itemize}[leftmargin=*, itemsep=0.2ex, parsep=0pt, topsep=0.2ex, partopsep=0pt]
  \item Treat ``fundamental mispricing'' as a safety cue; highly sensitive to over/undervaluation narratives.
  \item Feel uneasy about decoupling from fundamentals; doubt trend persistence as stretches grow.
  \item Read extremes as ``reversion is near,'' adopting cautious or even contrarian stances toward momentum.
  \item Stay skeptical of story-driven expansions; prioritize paths that reconnect cash flows, earnings, and valuation ranges.
  \item View price--value divergence as accumulating risk and seek early realignment.
\end{itemize}
\end{tcolorbox}

\begin{tcolorbox}[title=Price Misalignment Sensitivity (Weak / Inconsistent), fontupper=\footnotesize, breakable]
\textbf{Core belief:} Fundamental estimates are noisy and lagged; prices can deviate for long periods without immediate reversion.

\vspace{0.5mm}
\textbf{Behavioral tendencies:}
\begin{itemize}[leftmargin=*, itemsep=0.2ex, parsep=0pt, topsep=0.2ex, partopsep=0pt]
  \item Do not treat ``distance from fundamentals'' as a decisive trigger; let price information and market consensus speak first.
  \item Remain open to continuation---``expensive can get more expensive, cheap cheaper''---without forcing a reversion narrative.
  \item Treat ``mispricing'' as a soft signal unless stronger evidence emerges.
  \item Accept prolonged divergence without the burden of a must-revert anchor.
  \item Focus on verifiable present edge and risk management rather than intuitive ``ought-to-revert'' pulls.
\end{itemize}
\end{tcolorbox}

\captionof{figure}{Prompts — Price Misalignment Sensitivity}
\par\medskip

\subsection{Prompts for Trading Decisions}\label{prompt:trading}
We supply each agent with both fundamental and technical indicators together with daily stock data, using a unified prompt that standardizes how information is presented. The instruction explicitly encourages active trading by asking for a decisive BUY or SELL and reserving HOLD only when signals are uniformly neutral. This design can increase the volume and variability of executed trades, fostering a more dynamic, adaptable, and responsive trading environment that closely mirrors real-world market conditions.

\Needspace{18\baselineskip}

\begin{tcolorbox}[title=Trading Instruction for fundamental traders, fontupper=\footnotesize, breakable]
\textbf{Instruction:} Provide a decisive \textsc{buy} or \textsc{sell} unless indicators are uniformly neutral and \texttt{hold\_streak=0}. This study encourages active trading---prefer \textsc{buy}/\textsc{sell} and minimize \textsc{hold}.\\

\textbf{Persona Guidance:}\\
\{persona\_prompt\}\\

\textbf{Stock Daily Information and Fundametal Indicator:}\\
Date: \{date\}\\
CurrentRatio: \{current\_ratio\}\\
Leverage: \{leverage\}\\
FreeCashFlow: \{free\_cash\_flow\}\\
FCF\_to\_Capex: \{fcf\_to\_capex\}\\
Current\_Position: \{position\}\\
Return\_1d: \{ret1d\}\\
\end{tcolorbox}

\begin{tcolorbox}[title=Trading Instruction for technical traders, fontupper=\footnotesize, breakable]
\textbf{Instruction:} Provide a decisive \textsc{buy} or \textsc{sell} unless indicators are uniformly neutral and \texttt{hold\_streak=0}. This study encourages active trading---prefer \textsc{buy}/\textsc{sell} and minimize \textsc{hold}.\\

\textbf{Persona Guidance:}\\
\{persona\_prompt\}\\

\textbf{Stock Daily Information and Technical Indicator:}\\
Date: \{date\}\\
Price: \{price\}\\
Pre\_close\_price:  \{Pre\_close\_price\}\\
Change: \{change\}\\
Pct\_chg: \{pct\_chg\}\\
MACD: \{macd\}\\
MACD\_Signal: \{macd\_signal\}\\
MACD\_Hist: \{macd\_hist\}\\
Volatility\_20: \{vol\}\\
VolumeTrend\_20: \{volume\_trend\}\\
Return\_1d: \{ret1d\}\\
Current\_Position: \{position\}\\
\end{tcolorbox}

\captionof{figure}{Prompts — Trading Instruction for Agents}
\par\medskip

\subsection{Prompts for Style Switching}\label{prompt:switching}
The style switching prompt summarizes the agent’s recent context and asks for a Switch or Stay decision with a brief rationale. The goal is to frame switching as an explicit profit-maximizing trade-off between recent evidence and the agent’s predispositions, producing transparent and auditable behavior while allowing us to attribute switches to identifiable drivers rather than incidental prompt effects.

\Needspace{18\baselineskip}

\begin{tcolorbox}[title=Switching Prompt, fontupper=\footnotesize, breakable]
\begin{itemize}[leftmargin=*, itemsep=0.2ex, parsep=0pt, topsep=0.2ex, partopsep=0pt]
  \item You are an investment agent currently following the trading style: \{current\_style\}. 
  \item Your personality traits: aligned / not aligned with loss aversion; aligned / not aligned with herding; aligned / not aligned with wealth differentiation; aligned / not aligned with mispricing sensitivity. 
  \item You currently hold multiple stocks, and for each stock, the number of shares, price, and company fundamentals are as follows: Stock A: Shares \{shares\_A\}, Price \{price\_A\}, Debt-to-Asset Ratio \{debt\_asset\_ratio\_A\}, Current Ratio \{current\_ratio\_A\}, Net Operating Cash Flow \{operating\_cf\_A\}, Free Cash Flow \{free\_cf\_A\}. Stock B: Shares \{shares\_B\}, Price \{price\_B\}, Debt-to-Asset Ratio \{debt\_asset\_ratio\_B\}, Current Ratio \{current\_ratio\_B\}, Net Operating Cash Flow \{operating\_cf\_B\}, Free Cash Flow \{free\_cf\_B\}. \dots{} (list other stocks similarly) 
  \item You also have remaining funds of \{current\_fund\}. 
  \item During the same period, the opposite style has generated an average profit of \{opposite\_profit\}. 
  \item In the market, \{num\_current\} agents follow your current style, while \{num\_opposite\} agents follow the opposite style. 
  \item Your goal is to maximize profit in the stock market. Considering your recent gains or losses, the relative profitability of your current and opposite styles, and your personality traits (alignment or misalignment with loss aversion, herding, wealth differentiation, and mispricing sensitivity), decide whether to switch to the opposite style and provide your reasoning.
\end{itemize}
\end{tcolorbox}

\captionof{figure}{Prompts — Switching Instruction for Agents}
\par\medskip

\section{Rationale-Centric Case Studies}
\label{sec:case_studies}

To complement our score-based evaluation of style-switching alignment, we report case studies selected from daily logs where agents provided
non-empty \texttt{reason} strings.These cases illustrate how technical vs. fundamental styles and behavioral drivers are expressed in natural-language justifications, consistent with our simulator’s design of periodic style review and rationale reporting, providing a certain level of interpretability.

\subsection{Case Study 1: Technical Overbought Narrative Triggers Aggressive De-risking}
\noindent\textbf{Setting.}
On 2024-01-11, a technical-style agent executes a sell order with confidence=0.85, which provides a representative example of indicator-driven de-risking.

\noindent\textbf{Rationale evidence.}
The agent explicitly cites an extreme RSI overbought signal:

\begin{tcolorbox}[title=Reason Evidence, fontupper=\footnotesize, breakable]
\begin{itemize}[leftmargin=*, itemsep=0.2ex, parsep=0pt, topsep=0.2ex, partopsep=0pt]
  \item The RSI is extremely high at 92.63, indicating overbought conditions. This suggests a potential reversal or pullback, while the MACD is showing strong momentum but closes the gap with its signal line, indicating weakening buying momentum. The mispricing indicates a slight overvaluation. Given these factors, a sell position is warranted to capitalize on potential profit taking.
\end{itemize}
\end{tcolorbox}

\captionof{figure}{The reason evidence of case 1.}

\noindent\textbf{Interpretation.}
This case illustrates a typical technical-style justification that maps salient momentum/overbought signals (RSI/MACD) to a substantial de-risking action.

\subsection{Case Study 2: Fundamental Liquidity-Risk Narrative Drives Exposure Reduction}
\label{subsec:cs2}

\noindent\textbf{Setting.} On 2024-01-16, a fundamental-style agent executes a sell order with confidence=0.6, which provides a representative example of
liquidity-risk-driven exposure reduction.

\noindent\textbf{Rationale evidence.}
The rationale foregrounds liquidity risk:

\begin{tcolorbox}[title=Reason Evidence, fontupper=\footnotesize, breakable]
\begin{itemize}[leftmargin=*, itemsep=0.2ex, parsep=0pt, topsep=0.2ex, partopsep=0pt]
  \item The current ratio (0.86) indicates a potential liquidity issue, and leverage (0.98) is quite high, suggesting that the company may be taking on too much debt relative to its assets. The free cash flow is zero, and the negative FCF to CapEx (-0.17) indicates poor cash generation capacity, meaning the company might struggle to finance future growth or pay off debts. Additionally, while there is a slight mispricing indication (0.0515), it is not strong enough to override the concerning financial metrics. Therefore, it is prudent to reduce exposure.
\end{itemize}
\end{tcolorbox}

\captionof{figure}{The reason evidence of case 2.}

\noindent\textbf{Interpretation.}
Unlike indicator-driven technical rationales, this explanation prioritizes a
balance-sheet liquidity signal as the principal justification for exposure reduction,
consistent with the fundamental input space (e.g., current ratio).

\subsection{Case Study: Wealth-Differentiation- and Herding-Conditioned Style Switching}
\label{subsec:cs_switch_wealth_herding}

\noindent\textbf{Setting.} On 2024-08-06, at a scheduled style review point, a wealth-differentiation-sensitive agent with strong wealth salience and strong herding switches from fundamental to technical. At the time of review, technical style is the majority in the population, with 25 agents adopting the technical style compared to 7 with the fundamental style. The counterfactual comparison indicates a positive wealth advantage for the opposite style, with a wealth gap of 8137.78.

\noindent\textbf{Rationale evidence.}
The switch rationale jointly cites relative style performance and population-level adoption:

\begin{tcolorbox}[title=Reason Evidence, fontupper=\footnotesize, breakable]
\begin{itemize}[leftmargin=*, itemsep=0.2ex, parsep=0pt, topsep=0.2ex, partopsep=0pt]
  \item The opposite style has outperformed the current style significantly, and the high number of agents following it suggests a potential for greater profit, aligning with the goal of maximizing returns.
\end{itemize}
\end{tcolorbox}

\captionof{figure}{The reason evidence of case 3.}

\noindent\textbf{Interpretation.}
This case aligns with the agent’s design assumptions. As a strong/consistent
wealth-differentiation-sensitive agent, it frames switching as a response to the
wealth gap between styles, explicitly linking the decision to return maximization.
As a strong herding agent, it further treats broad adoption of the opposite style
as an additional profit signal, yielding a switch justification that combines

\end{document}